\documentclass[preprint,notitlepage,floatfix,aps,prd,showpacs,nofootinbib,amsmath,amssymb,amsfonts,superscriptaddress]{revtex4-1}

\usepackage{bm}
\usepackage{graphicx}
\usepackage{physics}
\usepackage{dsfont}
\usepackage[normalem]{ulem}
\usepackage{hyperref,url}
\usepackage{mathrsfs}
\usepackage{calrsfs}
\usepackage{tikz}
\usetikzlibrary{calc, decorations.pathmorphing, arrows.meta, positioning, backgrounds, fit, shapes.geometric}

\usepackage{comment}
\usepackage{soul}

\usepackage[caption=false]{subfig}
\renewcommand{\vec}[1]{\mathbf{#1}}
\usepackage{varioref}
\usepackage[active]{srcltx}

\expandafter\ifx\csname package@font\endcsname\relax\else
\expandafter\expandafter
\expandafter\usepackage
\expandafter\expandafter
\expandafter{\csname package@font\endcsname}%
\fi
\hypersetup{
	colorlinks=true,       
	linkcolor=blue,          
	citecolor=blue,        
}

\usepackage[section]{placeins}

\usepackage{fancybox}
\labelformat{figure}{Fig.~#1}

\newcommand{\ba}{\begin{align}}
\newcommand{\ea}{\end{align}}
\def\dm {\mathrm{d}}

\begin{document}
\begin{titlepage}
\pagestyle{empty}
\baselineskip=21pt
\vspace{2cm}
\begin{center}
{\bf {\Large The Gravitational Spectral \emph{Radio Forest}: A Signature of Primordial Black Holes}}
\end{center}
\begin{center}
\vskip 0.2in

{\bf P. George Christopher\footnotemark[1], K. Hari\footnotemark[1]},  and 
{\bf S. Shankaranarayanan\footnotemark[2]}\\

{\it Department of Physics, Indian Institute of Technology Bombay, Mumbai 400076, India}  \\
{\tt Email: p.georgechristopher@iitb.ac.in, hari.k@iitb.ac.in, shanki@iitb.ac.in}\\
\footnotetext[1]{Equal contribution to this work}
\footnotetext[2]{Corresponding Author}
\end{center}

\vspace*{0.5cm}

\begin{abstract}
\end{abstract}
We propose a novel gravitational signature to detect Primordial Black Hole (PBH) dark matter by treating interstellar hydrogen as a quantum sensor for spacetime curvature. 
Focusing on H\,II regions, we demonstrate that the Riemann tidal tensor of an \emph{asteroid-mass} PBH induces a symmetric splitting of the $2P_{3/2}$ state in bound hydrogen atoms.
This relativistic effect redistributes  $9.9\,\mathrm{GHz}$ absorption line into a \emph{gravitational spectral radio forest} with a bandwidth $\sim 2\,\mathrm{GHz}$.
By accounting for active accretion of Hydrogen atoms and the resulting density-squared emission measure within the Bondi radius, we find a relatively enhanced absorption spectrum. This feature presents a \emph{concrete, high-contrast} target for upcoming radio-surveys to constrain PBH populations in the dark matter sector. 

\vspace*{2.0cm}

\begin{center}
{\bf Received first prize in Gravity Research Foundation essay competition 2026.}
\end{center}

\end{titlepage}

\baselineskip=18pt

Throughout the history of modern astrophysics, paradigm shifts have been inextricably linked to the observation of specific atomic spectral lines. In 1929, Edwin Hubble's measurement of redshifted Calcium H and K absorption lines provided the first definitive proof of the expanding Universe, validating the dynamic nature of spacetime in General Relativity~\cite{Hubble:1929ig,Hubble:1937}. Decades later, the discovery of the Lyman-$\alpha$ forest in quasar spectra revealed the \emph{Cosmic Web} ---proving that the seemingly empty void between galaxies is actually threaded by massive filaments of dark matter~\cite{Rauch:1998xn}. Similarly, the ubiquitous $H\alpha$ emission line unveiled the Warm Ionized Medium that dominates interstellar space~\cite{Reynolds:1990abc}, while the 21-cm hyperfine transition of neutral hydrogen fundamentally shifted our ability to map galactic kinematics and probe the pre-stellar Epoch of Reionization~\cite{Furlanetto:2006jb}. In each instance, atomic spectroscopy transformed our understanding of the cosmos. 

Today, cosmology faces a \emph{persistent enigma}: the quest to identify the nature of dark matter~\cite{Bertone:2016nfn}. This endeavor began in earnest when Rubin and Ford's observations of spiral galaxy rotation curves revealed that the visible universe is \emph{gravitationally leaky} aka  the ratio of dark-to-light matter is about a factor of ten~\cite{Rubin:1970zza,Rubin:1980zd}.
Stellar velocities remained inexplicably flat far beyond the luminous disk~\cite{Rubin:1980zd}, a phenomenon later solidified by 21-cm radio maps, pointing to massive, non-luminous dark matter halos enveloping galaxies~\cite{Furlanetto:2006jb}. While the particle physics community has long favored Weakly Interacting Massive Particles as dark matter, the enduring absence of detection in direct capture experiments and collider searches has forced a widening of the theoretical horizon \cite{Roszkowski:2017nbc,Billard:2021uyg,Kahn:2021ttr,PerezdelosHeros:2020qyt,Misiaszek:2023sxe}.

Among the candidates for this missing mass, Primordial Black Holes (PBHs) stand out as a uniquely compelling, \emph{purely gravitational solution}~\cite{Zeldovich:1967lct,Hawking:1971ei}. 
They occupy a specific theoretical category as the only dark matter candidate that does not require the introduction of new particle species~\cite{Carr:1974nx,Carr:1975qj,Nadezhin:1978aa}. Although their formation is assumed to require a local enhancement of the primordial power spectrum, typically sourced by inflationary dynamics, the resulting compact objects are composed of Standard Model particles that have undergone relativistic collapse~\cite{Carr:2020xqk,Carr:2021bzv,Carr:2026hot,Green:2020jor,Escriva:2022duf,Shanki:2026}. A fundamental property distinguishing PBH from their astrophysical counterparts is their intrinsic spin. Unlike stellar black holes, which are born from rotating cores and typically retain high dimensionless spins ($a_* \sim 0.7-0.9$), PBHs formed in the standard radiation-dominated early universe are theoretically predicted to have negligible rotation ($a_* \approx 0$) \cite{DeLuca:2019buf,Mirbabayi:2019uph}. As irrotational relics of the high-energy universe \cite{Carr:1974nx,Carr:1975qj,Nadezhin:1978aa}, they behave as \emph{Cold Dark Matter} --- non-relativistic, effectively collisionless, and interacting solely via gravity.

PBHs serve as a probe of physical regimes inaccessible to terrestrial particle accelerators. Their formation links the non-linear dynamics of GR to quantum fluctuations generated during inflation at super-horizon scales~\cite{Green:1997sz,Sasaki:2018dmp,Musco:2018rwt,Escriva:2019phb}. Consequently, the PBH mass spectrum provides a record of the primordial power spectrum and the thermal history of the Universe~\cite{Byrnes:2018clq,Young:2019yug,Carr:2020xqk}. However, direct identification of PBHs
has not been achieved yet. This brings us to a fundamental question: \ul{If PBHs constitute a significant fraction of dark matter, do they leave a distinct, albeit subtle, imprint on the baryonic matter they encounter?}

\emph{In this essay}, we explore the possibility that a spectroscopic signature of gravitational origin may arise in the form of a novel atomic effect. We develop a new framework for detecting the local presence and distribution of asteroid-mass PBHs ($10^{17} - 10^{23}\,\mathrm{g}$) by treating interstellar hydrogen gas as a \emph{quantum sensor} for spacetime geometry via the \emph{Parker-Pimentel effect}~\cite{Parker:1980hlc,Parker:1980kw,Parker:1982nk}.

To understand the magnitude of the tidal forces required to perturb quantum states, let us evaluate the Kretschmann scalar ($K$) corresponding to a Schwarzschild space-time. At the event horizon, Kretschmann scalar exhibits a steep inverse scaling with mass, i. e., $K \propto M^{-4}$. While stellar-mass black holes possess negligible horizon curvature ($K \sim 10^{-13}\,\mathrm{m}^{-4}$), asteroid-mass PBHs ($10^{17} - 10^{23}\,\mathrm{g}$) harbor enormous localized curvature ($K \sim 10^{51}\,\mathrm{m}^{-4}$)~\cite{Shankaranarayanan:2022wbx,Mandal:2025xuc}. In other words, the tidal forces of the asteroid mass PBHs allow us to probe these perturbations using atomic physics.


Just as external electric or magnetic fields lead to the Stark and Zeeman splittings of atomic energy levels~\cite{Feynman2006Vol3}, gravitational fields can induce analogous shifts in atomic spectra. Parker~\cite{Parker:1980hlc, Parker:1982nk} demonstrated that for a single-electron atom in a curved spacetime background, the flat-space Schr\"odinger (or Dirac) equation receives well-defined gravitational corrections. 

To formalize interstellar hydrogen as a quantum sensor we must establish the perturbative corrections to its energy levels when captured in the gravitational potential of an asteroid-mass PBH with mass $M$. As mentioned earlier, since these PBHs are predicted to possess negligible intrinsic spin, we can rigorously model the local spacetime as a non-rotating, spherically symmetric geometry. Consider a hydrogen atom in a circular orbit around a PBH, effectively modeling gravitational capture or accretion. In the local inertial frame of the atom's nucleus, this spacetime metric can be expanded using Fermi Normal Coordinates~\cite{Manasse:1963zz,Poisson:2009pwt}.
By solving the Dirac equation in a curved spacetime background and reducing it to an effective Schr\"odinger-type equation, the resulting Hamiltonian acquires corrections beyond the standard kinetic and electrostatic terms in the form of a leading-order gravitational contribution $\delta \hat{H}_{\mathrm{grav}}$, governed by the spacetime curvature tensors~\cite{Parker:1980hlc}.
The gravitational correction to the energy levels takes the form:
\begin{equation}
    E^{(1)} = A R_{\hat{0}\hat{0}} + B R + \sum_{\hat{i}=1}^{3} C^{\hat{i}\hat{i}} R_{\hat{0}\hat{i}\hat{0}\hat{i}}~,
    \label{eqn:Parker-energy}
\end{equation}
where $A$, $B$, and $C^{\hat{i}\hat{i}}$ are state-dependent atomic coefficients, $R$ is the Ricci scalar, and $R_{\hat{\mu}\hat{\nu}}$ and $R_{\hat{\mu}\hat{\nu}\hat{\rho}\hat{\sigma}}$ are the Ricci and Riemann tensor components, respectively, evaluated in the local inertial frame.
Equation~\eqref{eqn:Parker-energy} reveals a fundamental dichotomy in the atomic response to spacetime curvature. As explicitly derived by Parker~\cite{Parker:1980kw}, 
the first-order correction in $S_{1/2}$ ($2P_{1/2}$) states solely occurs due to Ricci scalar and Ricci tensor terms as only even (odd) parity terms survive.
In the specific case of Schwarzschild spacetime, the Ricci curvature vanishes entirely, leaving these states unshifted. In the case of realistic non-vacuum or regular black hole solutions generate non-zero Ricci curvature \cite{Parvez:2025wtq}, however, these shifts are often suppressed by a minimal length scale.

For the $2P_{3/2}$ state, Parker and Pimentel~\cite{Parker:1982nk} showed that the first-order correction contains both Ricci and Riemann terms as both odd and even parity terms contribute to the energy corrections.
Hence, $2P_{3/2}$ level is highly sensitive to the {\emph{tidal part of Riemann tensor} term ($C^{\hat{i}\hat{i}} R_{\hat{0}\hat{i}\hat{0}\hat{i}}$). As shown in \ref{fig:h-energy-level}, even in a pure vacuum Schwarzschild geometry where Ricci terms vanish, the tidal forces of the PBH strictly break the spatial degeneracy of the atom. This induces splitting of the $2P_{3/2}$ level into two distinct components, symmetrically shifted about the unperturbed Quantum Electrodynamics (QED) energy. Because this splitting scales with the PBH mass and the inverse cube of the orbital radius, it provides a direct, localized probe of these primordial compact objects. This is a \emph{distinct feature} which can not be mimicked by other effects, like Doppler shift or competing transition lines. Interestingly, we also show that we can constrain the primordial dark matter fraction \emph{without} requiring prior knowledge of the precise mass function peak.
\begin{figure}[htbp]
    \centering
    \includegraphics[width=0.8\linewidth]{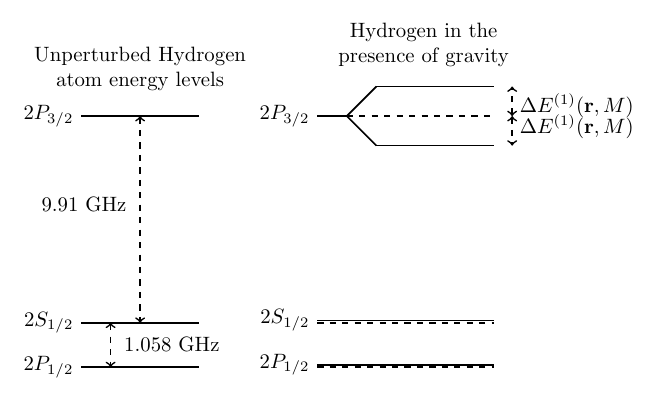}
    \caption{In the presence of gravity, the hydrogen $2S$ and $2P_{1/2}$ energy levels experience a baseline shift, whereas the $2P_{3/2}$ level is strictly broken by the Riemann tidal tensor, splitting into two distinct states (Gravitational fine splitting).}
    \label{fig:h-energy-level}
\end{figure}

This leads to the immediate question: \ul{What astrophysical environment can provide such a setup?} To observe this gravitational splitting, we require a massive population of hydrogen atoms in the $n=2$ excited state~\cite{Osterbrock:2006,Draine:2011}. As we demonstrate, H\,II regions --- vast clouds of ionized gas enveloping hot, young stars --- serve as the ideal laboratory~\cite{Anderson:2014,Wenger:2019,Guzman:2017xdg,Liu:2026}. 
These regions are characterized by intense ultraviolet radiation fields~\cite{Kulkarni:2022} that efficiently drive recombination processes and radiative cascades, continually pumping hydrogen into excited states. Crucially, the $2S_{1/2}$ level is metastable; its primary decay route to the ground state requires a highly suppressed two-photon emission. Consequently, the $2S_{1/2}$ state becomes highly overpopulated relative to the $2P$ states, creating a strong population inversion~\cite{Chatzikos:2023bkk}.

Because of this overpopulation, transitions between the $2S_{1/2}$ and $2P_{3/2}$ levels manifest in \emph{absorption} against the intense free-free (Bremsstrahlung) radio continuum naturally emitted by the nebula. As outlined by Dennison et al.~\cite{Dennison:2005fx}, standard QED~\cite{Bethe:1957ncq,CohenTannoudji:1977} dictates that atoms absorbing this local continuum radiation will transition from $2S_{1/2} \to 2P_{3/2}$ at a frequency of $\nu \approx 9.9\,\mathrm{GHz}~(3~\mathrm{cm})$.
\begin{figure}[!htbp]
    \centering
    \includegraphics[width=0.9\linewidth]{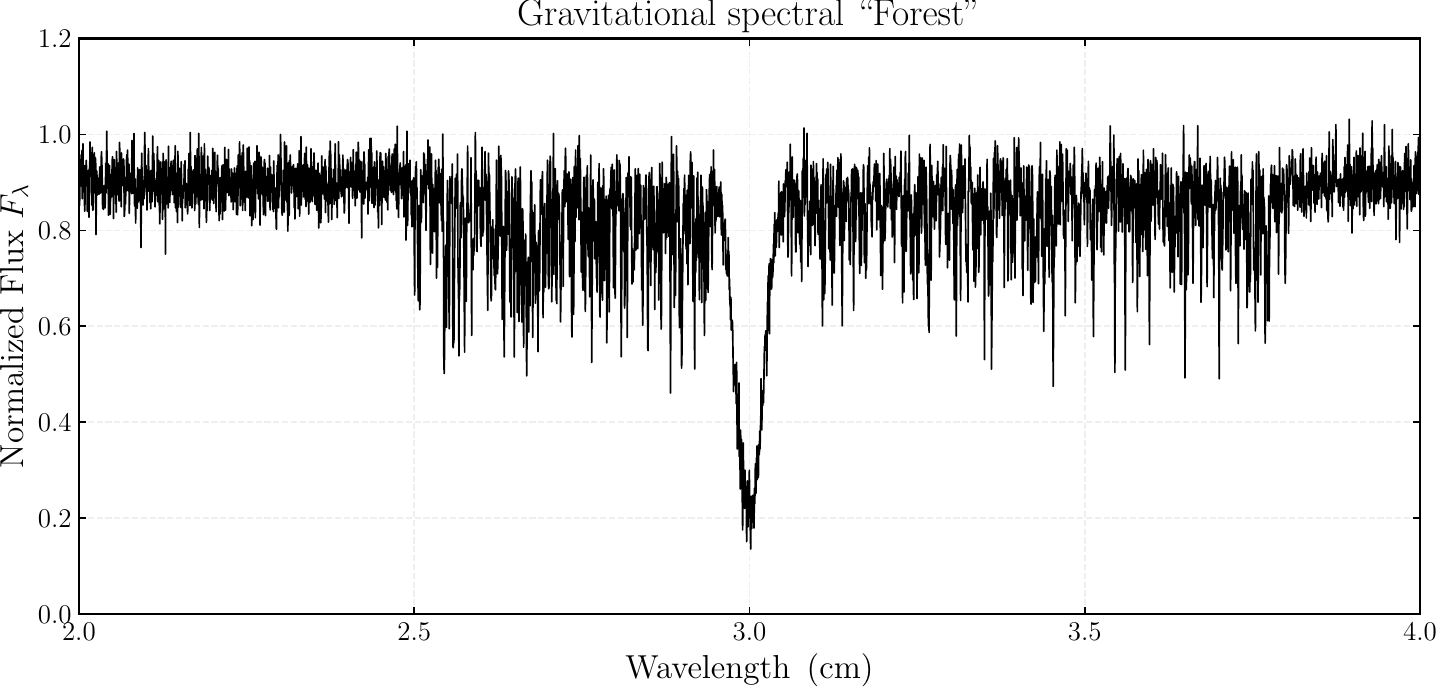}
    \caption{Illustration of the \textit{gravitational spectral forest} in radio observations, induced by the local Riemann curvature of PBHs in H\,II regions. The effect redistributes the central $9.9\,\mathrm{GHz}\;(3\,\mathrm{cm})$ transition into a broad, symmetric array of absorption sidebands spanning a $\sim 2\,\mathrm{GHz}$ bandwidth.}
    \label{fig:Gspectral-forest}
\end{figure}

While traditional observational cosmology expects to find a single, weak absorption trough at this frequency, our theoretical framework predicts a radically different signature in regions permeated by PBH dark matter. For the fraction of hydrogen atoms distributed in the immediate vicinity of an asteroid-mass PBH, the extreme localized curvature symmetrically splits the $2P_{3/2}$ energy level via the Parker-Pimentel effect (as shown in \ref{fig:h-energy-level}). 
As shown in \ref{fig:Gspectral-forest}, due to the splitting of the energy level, the absorption at $9.9\,\mathrm{GHz}$ transition predicted by Dennison et al.~\cite{Dennison:2005fx} will now be distributed to widely separated sidebands.
Over the statistical ensemble of PBH dark matter, this effect redistributes the single QED line into an expansive \emph{gravitational spectral radio forest} extending over a maximum bandwidth of $\sim 2\,\mathrm{GHz}$ in the radio regime.

To transition from a single theoretical atomic anomaly to a macroscopic observational signature, we must evaluate the statistical ensemble of the nebula. Within a vast H\,II region, the observed spectral profile will be a superposition of individual atomic shifts, dictated by both the spatial distribution of hydrogen atoms and the extended mass function of the PBH dark matter. This hierarchical framework is schematically illustrated in \ref{fig:model-schematic}.

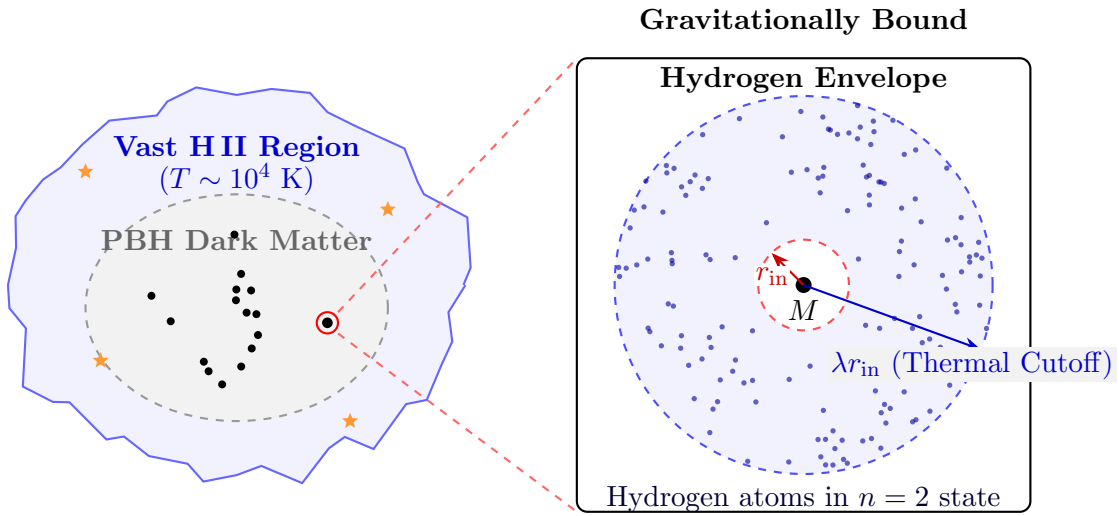
\begin{figure}[htbp]
    \centering
    \begin{tikzpicture}[x=1cm, y=1cm, >=Stealth]

        \begin{scope}[shift={(0,0)}]
            \draw[thick, color=blue!60, fill=blue!5, decorate, decoration={random steps,segment length=4mm,amplitude=2mm}] 
                (0,0) ellipse (3cm and 2.5cm);
            \node[color=blue!80!black, font=\bfseries] at (0, 1.8) {Vast H\,II Region};
            \node[color=blue!80!black, font=\small] at (0, 1.4) {($T \sim 10^4$ K)};

            \draw[thick, dashed, color=gray!80, fill=gray!10] (0,-0.3) ellipse (2cm and 1.5cm);
            \node[color=gray!80!black, font=\bfseries] at (0, 0.6) {PBH Dark Matter};

            \foreach \x/\y in {-2/1.5, 2/1, 1.5/-1.8, -1.8/-1} {
                \node[star, star points=5, star point ratio=2.25, fill=orange!80, inner sep=1pt] at (\x,\y) {};
            }
            \foreach \i in {1,...,15} {
                \pgfmathsetmacro{\ang}{rnd*360}
                \pgfmathsetmacro{\rad}{rnd*1.2}
                \fill[black] (\ang:\rad) ++(0,-0.3) circle (1.5pt);
            }

            \coordinate (ZoomPoint) at (1.2, -0.5);
            \fill[black] (ZoomPoint) circle (2pt);
            \draw[thick, red] (ZoomPoint) circle (4pt);
        \end{scope}

        \begin{scope}[shift={(7.5, 0)}]
            \draw[thick, rounded corners, fill=white] (-3, -3) rectangle (3, 3);
            
            \draw[thick, blue!70, dashed, fill=blue!5] (0,0) circle (2.5cm);
            
            \draw[thick, red!70, dashed, fill=white] (0,0) circle (0.6cm);

            \foreach \i in {1,...,150} {
                \pgfmathsetmacro{\ang}{rnd*360}
                \pgfmathsetmacro{\rad}{0.6 + sqrt(rnd)*(2.5-0.6)}
                \fill[blue!60!black, opacity=0.6] (\ang:\rad) circle (1pt);
            }

            \fill[black] (0,0) circle (3pt) node[below=2pt, font=\bfseries] {$M$};

            \draw[->, thick, red!80!black] (0,0) -- (135:0.6cm) 
                node[midway, below left=-2pt, font=\small, fill=white, inner sep=0.5pt] {$r_{\mathrm{in}}$};
            
            \draw[->, thick, blue!80!black] (0,0) -- (-20:2.5cm) 
                node[pos=0.95, below, font=\small, fill=black!5, inner sep=1pt] {$\lambda r_{\mathrm{in}}$ (Thermal Cutoff)};

            \node[font=\bfseries, align=center] at (0, 3.1) {Gravitationally Bound\\Hydrogen Envelope};
            \node[font=\small, align=center, black!80!blue] at (0, -2.83) {Hydrogen atoms in $n=2$ state};
        \end{scope}

        \draw[thick, dashed, red!60] (ZoomPoint) ++(4pt, 4pt) -- (4.5, 3);
        \draw[thick, dashed, red!60] (ZoomPoint) ++(4pt, -4pt) -- (4.5, -3);

    \end{tikzpicture}
    \caption{Schematic representation of the hierarchical statistical model. \textbf{Left:} The macroscopic astrophysical environment, where a vast H\,II region ($T \sim 10^4\,\mathrm{K}$) hosts a cosmological PBH dark matter halo. \textbf{Right:} A localized zoom-in on an individual asteroid-mass PBH. Each PBH captures a gravitationally bound envelope of excited hydrogen atoms, constrained strictly between an inner stable radius $r_{\mathrm{in}}$ and an outer thermal cutoff limit $\lambda r_{\mathrm{in}}$.}
    \label{fig:model-schematic}
\end{figure}
To model this, let the fraction of dark matter in the form of PBHs be $f_{\mathrm{PBH}}$. Then, the differential number density of PBHs of mass $M$ at a spatial position $\mathbf{r}$ can be written as:
\begin{equation}
    \frac{\dm n}{\dm M}(\vec{r},M) = f_{\mathrm{PBH}}\,\rho_{\mathrm{DM}}(\vec{r})\,\frac{\psi(M,t_0)}{M}\,,
\end{equation}
where $\rho_{\mathrm{DM}}(\vec{r})$ is the local dark matter density within the H\,II region, and $\psi(M,t_0)$ is the time evolved PBH mass function from the initial time to the current epoch $t_0$~\cite{Carr:2020xqk}. For the asteroid-mass range considered here, we saturate the dark matter abundance by setting $f_{\mathrm{PBH}} = 1$, implying that PBHs constitute the \emph{entirety of the dark matter density}. The probability of finding a PBH in the mass range $[M,M+\dm M]$ at position $\mathbf{r}$ within a volume element $\dm^3 \mathbf{r}$ is then given by:
\begin{equation}
p_{\mathrm{PBH}}(\mathbf{r},M)\,\dm^3\mathbf{r}\,\dm M = 
\frac{1}{N_{\mathrm{PBH}}} \left( \frac{\dm n}{\dm M} \right)\,\dm^3\mathbf{r}\,\dm M~,
\end{equation}
where $N_{\mathrm{PBH}} = \int\dm M\int\dm^3\mathbf{r}\,(\dm n/\dm M)$ is the total number of PBHs in the region. For the PBH mass function $\psi(M,t_0)$, we adopt the generalized critical collapse (GCC) model \cite{Klipfel:2026aug}. 

Next, we must model the localized hydrogen atmosphere around each PBH. \emph{To keep the calculations tractable}, we assume a uniform distribution of hydrogen atoms in circular orbits within a gravitationally bound envelope ranging from an inner radius $r_{\mathrm{in}}$ to an outer radius $r_{\mathrm{out}} = \lambda r_{\mathrm{in}}$.   
The inner radius is considered assuming the adiabatic conditions for the Parker's calculations are valid and the hydrogen atom is not distorted due to the tidal forces. 
This introduces a characteristic curvature length scale $D$ with a lower-bound $\sim 10^{-9}\, \mathrm{m}$
Crucially, the outer bound $\lambda$ is dictated by the thermal pressure 
of the H\,II region. In these environments, the typical gas temperature is $T \sim 10^4\,\mathrm{K}$. The hydrogen atmosphere cannot extend beyond the radius where the thermal kinetic energy exceeds the gravitational binding energy of the PBH ($G M m_p / r_{\mathrm{out}} \approx k_B T$). This dynamic competition restricts the cutoff parameter $(\lambda)$ for the asteroid mass range to be of the order of $100$. 
For the purpose of illustration, we first consider the case where all the hydrogen in the considered region is distributed within the spherical shell ($r_{\mathrm{in}},  \lambda r_{\mathrm{in}}$) --- \emph{Uniform shell model}. This defines our spatial probability density $p_{\mathrm{H}}(\mathbf{r}_c)$.

With both distributions established, the average energy correction for the macroscopic hydrogen population is given by convolving the atom's individual energy shift $E^{(1)}(\mathbf{r}_c,M)$ with these probabilities:
\begin{equation}
    \langle E^{(1)}\rangle =\int \dm^3 \mathbf{r} \int_{M_i}^{M_f} \dm M \int_{r_{\mathrm{in}}}^{\lambda r_{\mathrm{in}}} \dm^3 \mathbf{r}_c \;
    p_{\mathrm{PBH}}(\mathbf{r},M)\, p_{\mathrm{H}}(\mathbf{r}_c)\, E^{(1)}(\mathbf{r}_c,M)~.
    \label{eqn:avg-E-correction}
\end{equation}
Equation~\eqref{eqn:avg-E-correction} forms the central analytical result of our framework. It provides a direct, calculable bridge between a cosmological dark matter population and the resulting modification of atomic spectral lines.
To illustrate the expected signal, we evaluate this integral across the asteroid-mass window ($10^{17}\,\mathrm{g} \le M \le 10^{23}\,\mathrm{g}$), assuming characteristic curvature length scales in the range $10^{-7}\,\mathrm{m} \le D \le 10^{-2}\,\mathrm{m}$. At the lower bound, the available spatial volume limits the density of bound atoms; at the upper bound, the atoms cease to be gravitationally bound to the PBH mass range of interest. 

As shown in the contour plots of \ref{fig:Econtour-GHz}, substituting these astrophysical parameters yields a maximum statistical energy shift of order $\Delta \nu \sim 2\,\mathrm{GHz}$. While black holes of all masses theoretically induce a splitting of the $2P_{3/2}$ energy level, the extreme tidal curvature specific to the asteroid-mass regime produces a frequency shift on the order of $1 - 2\,\mathrm{GHz}$, placing the resulting signal squarely within the window of ground-based radio telescopes.
%
This creates an expansive, set of  symmetrically displaced absorption features. 
The resulting structure is a \textit{gravitational spectral radio forest} (as illustrated in \ref{fig:Gspectral-forest}). This forest of sidebands provides a distinct, and inherently \emph{gravitational fingerprint} for asteroid-mass dark matter.

\begin{figure}[htbp]
    \centering
    \includegraphics[width=0.49\linewidth]{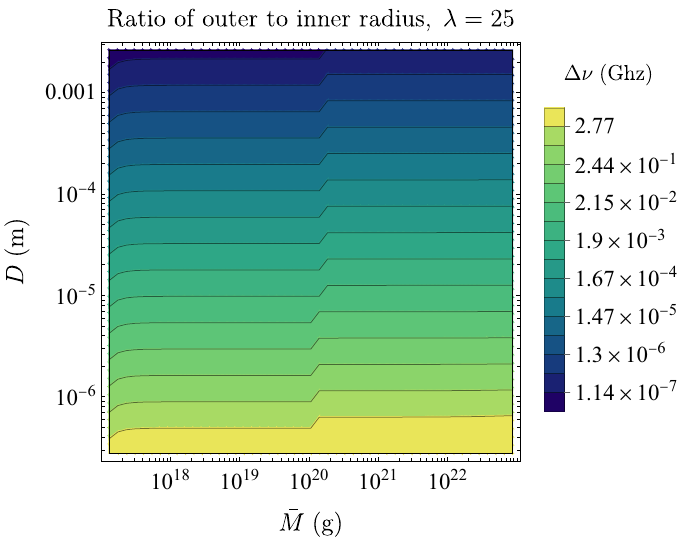}
    \includegraphics[width=0.49\linewidth]{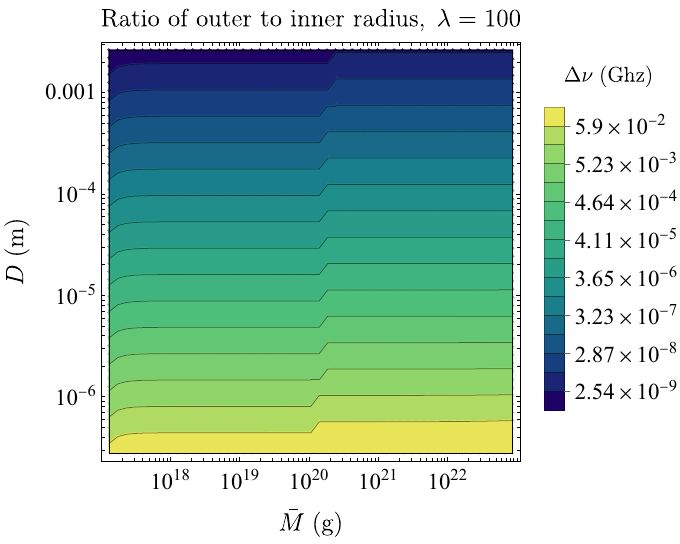}
    \caption{Contour plots illustrating the statistical mean frequency shift of the $2P_{3/2}$ hydrogen absorption line induced by a PBH distribution. The parameter space spans the PBH peak mass $\bar{M}$ and the curvature length scale $D$, evaluated for thermal cutoff parameters $\lambda = 25$ (\textbf{Left}) and $\lambda = 100$ (\textbf{Right}).}
    \label{fig:Econtour-GHz}
\end{figure}

{This leads to a fundamental observational question:} \ul{Can this gravitational spectral forest be translated into a direct signature for radio astronomy?} To answer this, we must evaluate the amplitude of the signal. In standard observations of H\,II regions, the strength of an absorption feature is quantified by the line-to-continuum ratio. The depth of our predicted gravitational sidebands, and the corresponding depletion of the central $9.9\,\mathrm{GHz}$ ($\sim 3\,\mathrm{cm}$) line, is strictly proportional to the fraction of the total hydrogen population that is gravitationally bound within the PBH regions. 

To evaluate this observational viability, let us go back to the uniform shell model and evaluate the number of hydrogen accreted in the shell. Taking into account the average $\rho_{\rm DM}$ from the observations, we find that the total number of hydrogen atoms bound in the shell is small and hence will not lead to a detectable signal. However, the simple model assumes that the PBH is a passive object and ignores the accretion around the PBH. 

Hence, we must move beyond a simple volume average and account for the fluid dynamics of the plasma around PBH, while ensuring that the energy corrections are of \emph{the same order} like in the above model. The sphere of gravitational influence is bounded by the Bondi radius, $r_B \approx 2GM/c_s^2$, where the gravitational potential equals the thermal kinetic energy ($c_s$ being the ambient sound speed)~\cite{bondi1952MNRAS}. (See Refs.~\cite{Franketal-Book,Armitage:2020owb} for a review on spherically symmetric, steady-state accretion flow.) This naturally fixes our thermal cutoff parameter, $\lambda r_{\mathrm{in}} \equiv r_B$.

Inside this radius, mass conservation dictates that the local hydrogen number density $n_{\mathrm{H}}^{\mathrm{(loc)}}(r)$ follows a steep free-fall accretion profile~\cite{bondi1952MNRAS}:
\begin{equation}
    n_{\mathrm{H}}^{\mathrm{(loc)}}(r) = n_{\mathrm{H}}^{\mathrm{(bg)}} \left(\frac{r_B}{r} \right)^{3/2} \quad \text{for } r_{\mathrm{in}} < r \le r_B~,
\end{equation}
where $n_{\mathrm{H}}^{\mathrm{(bg)}}$ is the ambient background density of the H\,II region. 
Crucially, the strength of the radio absorption feature --- quantified by the absorption coefficient $\kappa_\nu$ --- is governed by the recombination rate, which scales with the \emph{square} of the local density ($\propto n_e n_i \approx n_{\mathrm{H}}^2$). Therefore, the line-to-continuum ratio observed by a radio telescope is not strictly proportional to the raw number of bound atoms, but rather to the volume Emission Measure ($\mathrm{EM} = \int n^2 \dm^3x$). 

Since, the density square is proportional to $r^{-3}$, the emission measure integral yields a logarithmic divergence, $\ln(r_B / r_{\mathrm{in}})$. This implies that the overwhelming majority of the absorption signal originates from the extreme high-density shell sitting just outside $r_{\mathrm{in}}$. We define $r_{\mathrm{in}}$ as the \emph{collisional quenching radius}: the inner boundary where the gas density becomes so extreme that collisional transitions ($C_{sp} n_i$) overwhelm the radiative two-photon decay, physically destroying the metastable $2S_{1/2}$ population required for the $9.9\,\mathrm{GHz}$ transition. 

This quadratic ($n^2$) enhancement radically alters the observational prospects. If one evaluates the signal strictly in the asteroid-mass window ($M \sim 10^{19}\,\mathrm{g}$), the Bondi radius evaluates to a microscopic $\sim 1\,\mathrm{cm}$. 
While the individual Bondi volume of a $10^{19}\,\mathrm{g}$ PBH is microscopic ($\sim 1\,\mathrm{cm}$), this is counterbalanced by their sheer abundance. Since the PBH number density strictly scales as $n_{\mathrm{PBH}} \propto f_{\mathrm{PBH}}\rho_{\mathrm{DM}}/M$ \cite{Carr:2020xqk, Carr:2021bzv}, asteroid-mass PBHs are $\sim 10^{14}$ times more abundant than their solar-mass counterparts. This vast number density, combined with the $n^2$ enhancement at $r_{\mathrm{in}}$, integrates to a localized optical depth that can potentially lead to absorption on the Bremsstrahlung continuum.

Furthermore, integrating this accretion-assisted optical depth across the PBH mass spectrum reveals a striking analytical result. The total fractional absorption depends linearly on the local dark matter density $\rho_{\mathrm{DM}}(\mathbf{r})$ and the overall primordial fraction $f_{\mathrm{PBH}}$, but the explicit mass dependence effectively cancels out (as the extended bound volume scales proportionally with mass, while the PBH number density scales inversely). 
Consequently, measuring the line-to-continuum ratio of the $9.9\,\mathrm{GHz}$ spectral forest breaks a fundamental degeneracy. It allows radio observatories to constrain the primordial dark matter fraction $f_{\mathrm{PBH}}$ \emph{without} requiring prior knowledge of the precise mass function peak $\bar{M}$.

To conclude, the theoretical framework presented here establishes \emph{a direct link} between the perturbative relativistic quantum theory of atomic hydrogen in curved spacetime and macroscopic observables in astrophysical plasmas. By treating the extreme environment of H\,II regions as a laboratory, we have shown that the Riemann tidal tensor of PBHs induces a symmetric splitting of the metastable $2P_{3/2}$ state. Rather than a single weak absorption line at $9.9\,\mathrm{GHz}$, the presence of an asteroid PBH dark matter redistributes this transition into a \emph{gravitational spectral forest} spanning a $\sim 2\,\mathrm{GHz}$ bandwidth. While this essay focuses on the conceptual framework and macroscopic observables, to study the realistic scenario and the prospects of detection, an exhaustive analytical derivation of the relativistic atomic corrections, along with competing Rydberg lines, using 3D numerical simulations, will be presented in a forthcoming publication.

Since the amplitude of this signal isolates the dark matter fraction $f_{\mathrm{PBH}}$ from the mass function peak, it provides a highly constrained target for upcoming radio surveys. Detecting this  signal requires extreme baseline integration targeting hyper-dense H\,II regions (e.g., near the Galactic Center where the dark matter density profiles spike).
Ultimately, the subtle shifting of atomic energy levels may provide the definitive signature required to \emph{illuminate the dark, primordial universe}. The next decade can be an era of Gravitational Radio Astronomy.

\vspace{0.2in}
\noindent {\bf Acknowledgments} The authors thank Rekhesh Mohan for discussions. The authors also thank S.~Bhaumik, I.~Chakraborty, S.~Mahesh Chandran, A.~Kushwaha, Susmita Jana and Rekhesh Mohan for comments on the earlier draft. The work is supported by the ANRF Advanced Research grant (ANRF/ARG/2025/001514/PS). The MHRD fellowship at IIT Bombay financially supports PGC.
\vspace{0.2in}


\begin{thebibliography}{56}%
\makeatletter
\providecommand \@ifxundefined [1]{%
 \@ifx{#1\undefined}
}%
\providecommand \@ifnum [1]{%
 \ifnum #1\expandafter \@firstoftwo
 \else \expandafter \@secondoftwo
 \fi
}%
\providecommand \@ifx [1]{%
 \ifx #1\expandafter \@firstoftwo
 \else \expandafter \@secondoftwo
 \fi
}%
\providecommand \natexlab [1]{#1}%
\providecommand \enquote  [1]{``#1''}%
\providecommand \bibnamefont  [1]{#1}%
\providecommand \bibfnamefont [1]{#1}%
\providecommand \citenamefont [1]{#1}%
\providecommand \href@noop [0]{\@secondoftwo}%
\providecommand \href [0]{\begingroup \@sanitize@url \@href}%
\providecommand \@href[1]{\@@startlink{#1}\@@href}%
\providecommand \@@href[1]{\endgroup#1\@@endlink}%
\providecommand \@sanitize@url [0]{\catcode `\\12\catcode `\$12\catcode
  `\&12\catcode `\#12\catcode `\^12\catcode `\_12\catcode `\%12\relax}%
\providecommand \@@startlink[1]{}%
\providecommand \@@endlink[0]{}%
\providecommand \url  [0]{\begingroup\@sanitize@url \@url }%
\providecommand \@url [1]{\endgroup\@href {#1}{\urlprefix }}%
\providecommand \urlprefix  [0]{URL }%
\providecommand \Eprint [0]{\href }%
\providecommand \doibase [0]{http://dx.doi.org/}%
\providecommand \selectlanguage [0]{\@gobble}%
\providecommand \bibinfo  [0]{\@secondoftwo}%
\providecommand \bibfield  [0]{\@secondoftwo}%
\providecommand \translation [1]{[#1]}%
\providecommand \BibitemOpen [0]{}%
\providecommand \bibitemStop [0]{}%
\providecommand \bibitemNoStop [0]{.\EOS\space}%
\providecommand \EOS [0]{\spacefactor3000\relax}%
\providecommand \BibitemShut  [1]{\csname bibitem#1\endcsname}%
\let\auto@bib@innerbib\@empty
\bibitem [{\citenamefont {Hubble}(1929)}]{Hubble:1929ig}%
  \BibitemOpen
  \bibfield  {author} {\bibinfo {author} {\bibfnamefont {E.}~\bibnamefont
  {Hubble}},\ }\href {\doibase 10.1073/pnas.15.3.168} {\bibfield  {journal}
  {\bibinfo  {journal} {Proc. Nat. Acad. Sci.}\ }\textbf {\bibinfo {volume}
  {15}},\ \bibinfo {pages} {168} (\bibinfo {year} {1929})}\BibitemShut
  {NoStop}%
\bibitem [{\citenamefont {Hubble}(1937)}]{Hubble:1937}%
  \BibitemOpen
  \bibfield  {author} {\bibinfo {author} {\bibfnamefont {E.}~\bibnamefont
  {Hubble}},\ }\href
  {https://ned.ipac.caltech.edu/level5/Sept04/Hubble/paper.pdf} {\emph
  {\bibinfo {title} {The Observational Approach to Cosmology}}}\ (\bibinfo
  {publisher} {Oxford University Press, Clarendon Press},\ \bibinfo {address}
  {Oxford},\ \bibinfo {year} {1937})\ \bibinfo {note} {rhodes Memorial
  Lectures}\BibitemShut {NoStop}%
\bibitem [{\citenamefont {Rauch}(1998)}]{Rauch:1998xn}%
  \BibitemOpen
  \bibfield  {author} {\bibinfo {author} {\bibfnamefont {M.}~\bibnamefont
  {Rauch}},\ }\href {\doibase 10.1146/annurev.astro.36.1.267} {\bibfield
  {journal} {\bibinfo  {journal} {Ann. Rev. Astron. Astrophys.}\ }\textbf
  {\bibinfo {volume} {36}},\ \bibinfo {pages} {267} (\bibinfo {year} {1998})},\
  \Eprint {http://arxiv.org/abs/astro-ph/9806286} {arXiv:astro-ph/9806286}
  \BibitemShut {NoStop}%
\bibitem [{\citenamefont {{Reynolds}}(1990)}]{Reynolds:1990abc}%
  \BibitemOpen
  \bibfield  {author} {\bibinfo {author} {\bibfnamefont {R.~J.}\ \bibnamefont
  {{Reynolds}}},\ }in\ \href {\doibase
  10.1007/3-540-52891-110.1007/3-540-52891-1_116} {\emph {\bibinfo {booktitle}
  {Low Frequency Astrophysics from Space}}},\ Vol.\ \bibinfo {volume} {362},\
  \bibinfo {editor} {edited by\ \bibinfo {editor} {\bibfnamefont {N.~E.}\
  \bibnamefont {{Kassim}}}\ and\ \bibinfo {editor} {\bibfnamefont {K.~W.}\
  \bibnamefont {{Weiler}}}}\ (\bibinfo {year} {1990})\ p.\ \bibinfo {pages}
  {121}\BibitemShut {NoStop}%
\bibitem [{\citenamefont {Furlanetto}\ \emph {et~al.}(2006)\citenamefont
  {Furlanetto}, \citenamefont {Oh},\ and\ \citenamefont
  {Briggs}}]{Furlanetto:2006jb}%
  \BibitemOpen
  \bibfield  {author} {\bibinfo {author} {\bibfnamefont {S.}~\bibnamefont
  {Furlanetto}}, \bibinfo {author} {\bibfnamefont {S.~P.}\ \bibnamefont {Oh}},
  \ and\ \bibinfo {author} {\bibfnamefont {F.}~\bibnamefont {Briggs}},\ }\href
  {\doibase 10.1016/j.physrep.2006.08.002} {\bibfield  {journal} {\bibinfo
  {journal} {Phys. Rept.}\ }\textbf {\bibinfo {volume} {433}},\ \bibinfo
  {pages} {181} (\bibinfo {year} {2006})},\ \Eprint
  {http://arxiv.org/abs/astro-ph/0608032} {arXiv:astro-ph/0608032} \BibitemShut
  {NoStop}%
\bibitem [{\citenamefont {Bertone}\ and\ \citenamefont
  {Hooper}(2018)}]{Bertone:2016nfn}%
  \BibitemOpen
  \bibfield  {author} {\bibinfo {author} {\bibfnamefont {G.}~\bibnamefont
  {Bertone}}\ and\ \bibinfo {author} {\bibfnamefont {D.}~\bibnamefont
  {Hooper}},\ }\href {\doibase 10.1103/RevModPhys.90.045002} {\bibfield
  {journal} {\bibinfo  {journal} {Rev. Mod. Phys.}\ }\textbf {\bibinfo {volume}
  {90}},\ \bibinfo {pages} {045002} (\bibinfo {year} {2018})},\ \Eprint
  {http://arxiv.org/abs/1605.04909} {arXiv:1605.04909 [astro-ph.CO]}
  \BibitemShut {NoStop}%
\bibitem [{\citenamefont {Rubin}\ and\ \citenamefont
  {Ford}(1970)}]{Rubin:1970zza}%
  \BibitemOpen
  \bibfield  {author} {\bibinfo {author} {\bibfnamefont {V.~C.}\ \bibnamefont
  {Rubin}}\ and\ \bibinfo {author} {\bibfnamefont {W.~K.}\ \bibnamefont {Ford},
  \bibfnamefont {Jr.}},\ }\href {\doibase 10.1086/150317} {\bibfield  {journal}
  {\bibinfo  {journal} {Astrophys. J.}\ }\textbf {\bibinfo {volume} {159}},\
  \bibinfo {pages} {379} (\bibinfo {year} {1970})}\BibitemShut {NoStop}%
\bibitem [{\citenamefont {Rubin}\ \emph {et~al.}(1980)\citenamefont {Rubin},
  \citenamefont {Thonnard},\ and\ \citenamefont {Ford}}]{Rubin:1980zd}%
  \BibitemOpen
  \bibfield  {author} {\bibinfo {author} {\bibfnamefont {V.~C.}\ \bibnamefont
  {Rubin}}, \bibinfo {author} {\bibfnamefont {N.}~\bibnamefont {Thonnard}}, \
  and\ \bibinfo {author} {\bibfnamefont {W.~K.}\ \bibnamefont {Ford},
  \bibfnamefont {Jr.}},\ }\href {\doibase 10.1086/158003} {\bibfield  {journal}
  {\bibinfo  {journal} {Astrophys. J.}\ }\textbf {\bibinfo {volume} {238}},\
  \bibinfo {pages} {471} (\bibinfo {year} {1980})}\BibitemShut {NoStop}%
\bibitem [{\citenamefont {Roszkowski}\ \emph {et~al.}(2018)\citenamefont
  {Roszkowski}, \citenamefont {Sessolo},\ and\ \citenamefont
  {Trojanowski}}]{Roszkowski:2017nbc}%
  \BibitemOpen
  \bibfield  {author} {\bibinfo {author} {\bibfnamefont {L.}~\bibnamefont
  {Roszkowski}}, \bibinfo {author} {\bibfnamefont {E.~M.}\ \bibnamefont
  {Sessolo}}, \ and\ \bibinfo {author} {\bibfnamefont {S.}~\bibnamefont
  {Trojanowski}},\ }\href {\doibase 10.1088/1361-6633/aab913} {\bibfield
  {journal} {\bibinfo  {journal} {Rept. Prog. Phys.}\ }\textbf {\bibinfo
  {volume} {81}},\ \bibinfo {pages} {066201} (\bibinfo {year} {2018})},\
  \Eprint {http://arxiv.org/abs/1707.06277} {arXiv:1707.06277 [hep-ph]}
  \BibitemShut {NoStop}%
\bibitem [{\citenamefont {Billard}\ \emph {et~al.}(2022)\citenamefont {Billard}
  \emph {et~al.}}]{Billard:2021uyg}%
  \BibitemOpen
  \bibfield  {author} {\bibinfo {author} {\bibfnamefont {J.}~\bibnamefont
  {Billard}} \emph {et~al.},\ }\href {\doibase 10.1088/1361-6633/ac5754}
  {\bibfield  {journal} {\bibinfo  {journal} {Rept. Prog. Phys.}\ }\textbf
  {\bibinfo {volume} {85}},\ \bibinfo {pages} {056201} (\bibinfo {year}
  {2022})},\ \Eprint {http://arxiv.org/abs/2104.07634} {arXiv:2104.07634
  [hep-ex]} \BibitemShut {NoStop}%
\bibitem [{\citenamefont {Kahn}\ and\ \citenamefont
  {Lin}(2022)}]{Kahn:2021ttr}%
  \BibitemOpen
  \bibfield  {author} {\bibinfo {author} {\bibfnamefont {Y.}~\bibnamefont
  {Kahn}}\ and\ \bibinfo {author} {\bibfnamefont {T.}~\bibnamefont {Lin}},\
  }\href {\doibase 10.1088/1361-6633/ac5f63} {\bibfield  {journal} {\bibinfo
  {journal} {Rept. Prog. Phys.}\ }\textbf {\bibinfo {volume} {85}},\ \bibinfo
  {pages} {066901} (\bibinfo {year} {2022})},\ \Eprint
  {http://arxiv.org/abs/2108.03239} {arXiv:2108.03239 [hep-ph]} \BibitemShut
  {NoStop}%
\bibitem [{\citenamefont {P{\'e}rez de~los
  Heros}(2020)}]{PerezdelosHeros:2020qyt}%
  \BibitemOpen
  \bibfield  {author} {\bibinfo {author} {\bibfnamefont {C.}~\bibnamefont
  {P{\'e}rez de~los Heros}},\ }\href {\doibase 10.3390/sym12101648} {\bibfield
  {journal} {\bibinfo  {journal} {Symmetry}\ }\textbf {\bibinfo {volume}
  {12}},\ \bibinfo {pages} {1648} (\bibinfo {year} {2020})},\ \Eprint
  {http://arxiv.org/abs/2008.11561} {arXiv:2008.11561 [astro-ph.HE]}
  \BibitemShut {NoStop}%
\bibitem [{\citenamefont {Misiaszek}\ and\ \citenamefont
  {Rossi}(2024)}]{Misiaszek:2023sxe}%
  \BibitemOpen
  \bibfield  {author} {\bibinfo {author} {\bibfnamefont {M.}~\bibnamefont
  {Misiaszek}}\ and\ \bibinfo {author} {\bibfnamefont {N.}~\bibnamefont
  {Rossi}},\ }\href {\doibase 10.3390/sym16020201} {\bibfield  {journal}
  {\bibinfo  {journal} {Symmetry}\ }\textbf {\bibinfo {volume} {16}},\ \bibinfo
  {pages} {201} (\bibinfo {year} {2024})},\ \Eprint
  {http://arxiv.org/abs/2310.20472} {arXiv:2310.20472 [hep-ph]} \BibitemShut
  {NoStop}%
\bibitem [{\citenamefont {Zel'dovich}\ and\ \citenamefont
  {Novikov}(1967)}]{Zeldovich:1967lct}%
  \BibitemOpen
  \bibfield  {author} {\bibinfo {author} {\bibfnamefont {Y.~B.}\ \bibnamefont
  {Zel'dovich}}\ and\ \bibinfo {author} {\bibfnamefont {I.~D.}\ \bibnamefont
  {Novikov}},\ }\href@noop {} {\bibfield  {journal} {\bibinfo  {journal} {Sov.
  Astron.}\ }\textbf {\bibinfo {volume} {10}},\ \bibinfo {pages} {602}
  (\bibinfo {year} {1967})}\BibitemShut {NoStop}%
\bibitem [{\citenamefont {Hawking}(1971)}]{Hawking:1971ei}%
  \BibitemOpen
  \bibfield  {author} {\bibinfo {author} {\bibfnamefont {S.}~\bibnamefont
  {Hawking}},\ }\href {\doibase 10.1093/mnras/152.1.75} {\bibfield  {journal}
  {\bibinfo  {journal} {Mon. Not. Roy. Astron. Soc.}\ }\textbf {\bibinfo
  {volume} {152}},\ \bibinfo {pages} {75} (\bibinfo {year} {1971})}\BibitemShut
  {NoStop}%
\bibitem [{\citenamefont {Carr}\ and\ \citenamefont
  {Hawking}(1974)}]{Carr:1974nx}%
  \BibitemOpen
  \bibfield  {author} {\bibinfo {author} {\bibfnamefont {B.~J.}\ \bibnamefont
  {Carr}}\ and\ \bibinfo {author} {\bibfnamefont {S.~W.}\ \bibnamefont
  {Hawking}},\ }\href {\doibase 10.1093/mnras/168.2.399} {\bibfield  {journal}
  {\bibinfo  {journal} {Mon. Not. Roy. Astron. Soc.}\ }\textbf {\bibinfo
  {volume} {168}},\ \bibinfo {pages} {399} (\bibinfo {year}
  {1974})}\BibitemShut {NoStop}%
\bibitem [{\citenamefont {Carr}(1975)}]{Carr:1975qj}%
  \BibitemOpen
  \bibfield  {author} {\bibinfo {author} {\bibfnamefont {B.~J.}\ \bibnamefont
  {Carr}},\ }\href {\doibase 10.1086/153853} {\bibfield  {journal} {\bibinfo
  {journal} {Astrophys. J.}\ }\textbf {\bibinfo {volume} {201}},\ \bibinfo
  {pages} {1} (\bibinfo {year} {1975})}\BibitemShut {NoStop}%
\bibitem [{\citenamefont {{Nadezhin}}\ \emph {et~al.}(1978)\citenamefont
  {{Nadezhin}}, \citenamefont {{Novikov}},\ and\ \citenamefont
  {{Polnarev}}}]{Nadezhin:1978aa}%
  \BibitemOpen
  \bibfield  {author} {\bibinfo {author} {\bibfnamefont {D.~K.}\ \bibnamefont
  {{Nadezhin}}}, \bibinfo {author} {\bibfnamefont {I.~D.}\ \bibnamefont
  {{Novikov}}}, \ and\ \bibinfo {author} {\bibfnamefont {A.~G.}\ \bibnamefont
  {{Polnarev}}},\ }\href@noop {} {\bibfield  {journal} {\bibinfo  {journal}
  {Sov. Astronomy}\ }\textbf {\bibinfo {volume} {22}},\ \bibinfo {pages} {129}
  (\bibinfo {year} {1978})}\BibitemShut {NoStop}%
\bibitem [{\citenamefont {Carr}\ \emph {et~al.}(2021)\citenamefont {Carr},
  \citenamefont {Kohri}, \citenamefont {Sendouda},\ and\ \citenamefont
  {Yokoyama}}]{Carr:2020xqk}%
  \BibitemOpen
  \bibfield  {author} {\bibinfo {author} {\bibfnamefont {B.}~\bibnamefont
  {Carr}}, \bibinfo {author} {\bibfnamefont {K.}~\bibnamefont {Kohri}},
  \bibinfo {author} {\bibfnamefont {Y.}~\bibnamefont {Sendouda}}, \ and\
  \bibinfo {author} {\bibfnamefont {J.}~\bibnamefont {Yokoyama}},\ }\href@noop
  {} {\bibfield  {journal} {\bibinfo  {journal} {Rept. Prog. Phys.}\ }\textbf
  {\bibinfo {volume} {84}},\ \bibinfo {pages} {116902} (\bibinfo {year}
  {2021})},\ \Eprint {http://arxiv.org/abs/2002.12778} {arXiv:2002.12778
  [astro-ph.CO]} \BibitemShut {NoStop}%
\bibitem [{\citenamefont {Carr}\ and\ \citenamefont
  {Kuhnel}(2022)}]{Carr:2021bzv}%
  \BibitemOpen
  \bibfield  {author} {\bibinfo {author} {\bibfnamefont {B.}~\bibnamefont
  {Carr}}\ and\ \bibinfo {author} {\bibfnamefont {F.}~\bibnamefont {Kuhnel}},\
  }\href {\doibase 10.21468/SciPostPhysLectNotes.48} {\bibfield  {journal}
  {\bibinfo  {journal} {SciPost Phys. Lect. Notes}\ }\textbf {\bibinfo {volume}
  {48}},\ \bibinfo {pages} {1} (\bibinfo {year} {2022})},\ \Eprint
  {http://arxiv.org/abs/2110.02821} {arXiv:2110.02821 [astro-ph.CO]}
  \BibitemShut {NoStop}%
\bibitem [{\citenamefont {Carr}\ \emph {et~al.}(2026)\citenamefont {Carr},
  \citenamefont {Iovino}, \citenamefont {Perna}, \citenamefont {Vaskonen},\
  and\ \citenamefont {Veerm{\"a}e}}]{Carr:2026hot}%
  \BibitemOpen
  \bibfield  {author} {\bibinfo {author} {\bibfnamefont {B.}~\bibnamefont
  {Carr}}, \bibinfo {author} {\bibfnamefont {A.~J.}\ \bibnamefont {Iovino}},
  \bibinfo {author} {\bibfnamefont {G.}~\bibnamefont {Perna}}, \bibinfo
  {author} {\bibfnamefont {V.}~\bibnamefont {Vaskonen}}, \ and\ \bibinfo
  {author} {\bibfnamefont {H.}~\bibnamefont {Veerm{\"a}e}},\ }\href@noop {}
  {\enquote {\bibinfo {title} {{Primordial black holes: constraints, potential
  evidence and prospects}},}\ } (\bibinfo {year} {2026}),\ \Eprint
  {http://arxiv.org/abs/2601.06024} {arXiv:2601.06024 [astro-ph.CO]}
  \BibitemShut {NoStop}%
\bibitem [{\citenamefont {Green}\ and\ \citenamefont
  {Kavanagh}(2021)}]{Green:2020jor}%
  \BibitemOpen
  \bibfield  {author} {\bibinfo {author} {\bibfnamefont {A.~M.}\ \bibnamefont
  {Green}}\ and\ \bibinfo {author} {\bibfnamefont {B.~J.}\ \bibnamefont
  {Kavanagh}},\ }\href {\doibase 10.1088/1361-6471/abc534} {\bibfield
  {journal} {\bibinfo  {journal} {J. Phys. G}\ }\textbf {\bibinfo {volume}
  {48}},\ \bibinfo {pages} {043001} (\bibinfo {year} {2021})},\ \Eprint
  {http://arxiv.org/abs/2007.10722} {arXiv:2007.10722 [astro-ph.CO]}
  \BibitemShut {NoStop}%
\bibitem [{\citenamefont {Escriv{\`a}}\ \emph {et~al.}(2022)\citenamefont
  {Escriv{\`a}}, \citenamefont {Kuhnel},\ and\ \citenamefont
  {Tada}}]{Escriva:2022duf}%
  \BibitemOpen
  \bibfield  {author} {\bibinfo {author} {\bibfnamefont {A.}~\bibnamefont
  {Escriv{\`a}}}, \bibinfo {author} {\bibfnamefont {F.}~\bibnamefont {Kuhnel}},
  \ and\ \bibinfo {author} {\bibfnamefont {Y.}~\bibnamefont {Tada}},\ }\href
  {\doibase 10.1016/B978-0-32-395636-9.00012-8} {\bibfield  {journal} {\bibinfo
   {journal} {Arxiv}\ } (\bibinfo {year} {2022}),\
  10.1016/B978-0-32-395636-9.00012-8},\ \Eprint
  {http://arxiv.org/abs/2211.05767} {arXiv:2211.05767 [astro-ph.CO]}
  \BibitemShut {NoStop}%
\bibitem [{\citenamefont {Shankaranarayanan}\ \emph {et~al.}(2026)\citenamefont
  {Shankaranarayanan}, \citenamefont {Bhattacharya},\ and\ \citenamefont
  {Vidyarthi}}]{Shanki:2026}%
  \BibitemOpen
  \bibfield  {author} {\bibinfo {author} {\bibfnamefont {S.}~\bibnamefont
  {Shankaranarayanan}}, \bibinfo {author} {\bibfnamefont {S.}~\bibnamefont
  {Bhattacharya}}, \ and\ \bibinfo {author} {\bibfnamefont {A.}~\bibnamefont
  {Vidyarthi}},\ }\href@noop {} {\bibfield  {journal} {\bibinfo  {journal}
  {Invited Review for centenary issue of Indian Journal of Physics}\ }
  (\bibinfo {year} {2026})}\BibitemShut {NoStop}%
\bibitem [{\citenamefont {De~Luca}\ \emph {et~al.}(2019)\citenamefont
  {De~Luca}, \citenamefont {Desjacques}, \citenamefont {Franciolini},
  \citenamefont {Malhotra},\ and\ \citenamefont {Riotto}}]{DeLuca:2019buf}%
  \BibitemOpen
  \bibfield  {author} {\bibinfo {author} {\bibfnamefont {V.}~\bibnamefont
  {De~Luca}}, \bibinfo {author} {\bibfnamefont {V.}~\bibnamefont {Desjacques}},
  \bibinfo {author} {\bibfnamefont {G.}~\bibnamefont {Franciolini}}, \bibinfo
  {author} {\bibfnamefont {A.}~\bibnamefont {Malhotra}}, \ and\ \bibinfo
  {author} {\bibfnamefont {A.}~\bibnamefont {Riotto}},\ }\href {\doibase
  10.1088/1475-7516/2019/05/018} {\bibfield  {journal} {\bibinfo  {journal}
  {JCAP}\ }\textbf {\bibinfo {volume} {05}},\ \bibinfo {pages} {018} (\bibinfo
  {year} {2019})},\ \Eprint {http://arxiv.org/abs/1903.01179} {arXiv:1903.01179
  [astro-ph.CO]} \BibitemShut {NoStop}%
\bibitem [{\citenamefont {Mirbabayi}\ \emph {et~al.}(2020)\citenamefont
  {Mirbabayi}, \citenamefont {Gruzinov},\ and\ \citenamefont
  {Nore{\~n}a}}]{Mirbabayi:2019uph}%
  \BibitemOpen
  \bibfield  {author} {\bibinfo {author} {\bibfnamefont {M.}~\bibnamefont
  {Mirbabayi}}, \bibinfo {author} {\bibfnamefont {A.}~\bibnamefont {Gruzinov}},
  \ and\ \bibinfo {author} {\bibfnamefont {J.}~\bibnamefont {Nore{\~n}a}},\
  }\href {\doibase 10.1088/1475-7516/2020/03/017} {\bibfield  {journal}
  {\bibinfo  {journal} {JCAP}\ }\textbf {\bibinfo {volume} {03}},\ \bibinfo
  {pages} {017} (\bibinfo {year} {2020})},\ \Eprint
  {http://arxiv.org/abs/1901.05963} {arXiv:1901.05963 [astro-ph.CO]}
  \BibitemShut {NoStop}%
\bibitem [{\citenamefont {Green}\ and\ \citenamefont
  {Liddle}(1997)}]{Green:1997sz}%
  \BibitemOpen
  \bibfield  {author} {\bibinfo {author} {\bibfnamefont {A.~M.}\ \bibnamefont
  {Green}}\ and\ \bibinfo {author} {\bibfnamefont {A.~R.}\ \bibnamefont
  {Liddle}},\ }\href {\doibase 10.1103/PhysRevD.56.6166} {\bibfield  {journal}
  {\bibinfo  {journal} {Phys. Rev. D}\ }\textbf {\bibinfo {volume} {56}},\
  \bibinfo {pages} {6166} (\bibinfo {year} {1997})},\ \Eprint
  {http://arxiv.org/abs/astro-ph/9704251} {arXiv:astro-ph/9704251} \BibitemShut
  {NoStop}%
\bibitem [{\citenamefont {Sasaki}\ \emph {et~al.}(2018)\citenamefont {Sasaki},
  \citenamefont {Suyama}, \citenamefont {Tanaka},\ and\ \citenamefont
  {Yokoyama}}]{Sasaki:2018dmp}%
  \BibitemOpen
  \bibfield  {author} {\bibinfo {author} {\bibfnamefont {M.}~\bibnamefont
  {Sasaki}}, \bibinfo {author} {\bibfnamefont {T.}~\bibnamefont {Suyama}},
  \bibinfo {author} {\bibfnamefont {T.}~\bibnamefont {Tanaka}}, \ and\ \bibinfo
  {author} {\bibfnamefont {S.}~\bibnamefont {Yokoyama}},\ }\href {\doibase
  10.1088/1361-6382/aaa7b4} {\bibfield  {journal} {\bibinfo  {journal} {Class.
  Quant. Grav.}\ }\textbf {\bibinfo {volume} {35}},\ \bibinfo {pages} {063001}
  (\bibinfo {year} {2018})},\ \Eprint {http://arxiv.org/abs/1801.05235}
  {arXiv:1801.05235 [astro-ph.CO]} \BibitemShut {NoStop}%
\bibitem [{\citenamefont {Musco}(2019)}]{Musco:2018rwt}%
  \BibitemOpen
  \bibfield  {author} {\bibinfo {author} {\bibfnamefont {I.}~\bibnamefont
  {Musco}},\ }\href {\doibase 10.1103/PhysRevD.100.123524} {\bibfield
  {journal} {\bibinfo  {journal} {Phys. Rev. D}\ }\textbf {\bibinfo {volume}
  {100}},\ \bibinfo {pages} {123524} (\bibinfo {year} {2019})},\ \Eprint
  {http://arxiv.org/abs/1809.02127} {arXiv:1809.02127 [gr-qc]} \BibitemShut
  {NoStop}%
\bibitem [{\citenamefont {Escriv\`a}\ \emph {et~al.}(2020)\citenamefont
  {Escriv\`a}, \citenamefont {Germani},\ and\ \citenamefont
  {Sheth}}]{Escriva:2019phb}%
  \BibitemOpen
  \bibfield  {author} {\bibinfo {author} {\bibfnamefont {A.}~\bibnamefont
  {Escriv\`a}}, \bibinfo {author} {\bibfnamefont {C.}~\bibnamefont {Germani}},
  \ and\ \bibinfo {author} {\bibfnamefont {R.~K.}\ \bibnamefont {Sheth}},\
  }\href {\doibase 10.1103/PhysRevD.101.044022} {\bibfield  {journal} {\bibinfo
   {journal} {Phys. Rev. D}\ }\textbf {\bibinfo {volume} {101}},\ \bibinfo
  {pages} {044022} (\bibinfo {year} {2020})},\ \Eprint
  {http://arxiv.org/abs/1907.13311} {arXiv:1907.13311 [gr-qc]} \BibitemShut
  {NoStop}%
\bibitem [{\citenamefont {Byrnes}\ \emph {et~al.}(2018)\citenamefont {Byrnes},
  \citenamefont {Hindmarsh}, \citenamefont {Young},\ and\ \citenamefont
  {Hawkins}}]{Byrnes:2018clq}%
  \BibitemOpen
  \bibfield  {author} {\bibinfo {author} {\bibfnamefont {C.~T.}\ \bibnamefont
  {Byrnes}}, \bibinfo {author} {\bibfnamefont {M.}~\bibnamefont {Hindmarsh}},
  \bibinfo {author} {\bibfnamefont {S.}~\bibnamefont {Young}}, \ and\ \bibinfo
  {author} {\bibfnamefont {M.~R.~S.}\ \bibnamefont {Hawkins}},\ }\href
  {\doibase 10.1088/1475-7516/2018/08/041} {\bibfield  {journal} {\bibinfo
  {journal} {JCAP}\ }\textbf {\bibinfo {volume} {08}},\ \bibinfo {pages} {041}
  (\bibinfo {year} {2018})},\ \Eprint {http://arxiv.org/abs/1801.06138}
  {arXiv:1801.06138 [astro-ph.CO]} \BibitemShut {NoStop}%
\bibitem [{\citenamefont {Young}\ \emph {et~al.}(2019)\citenamefont {Young},
  \citenamefont {Musco},\ and\ \citenamefont {Byrnes}}]{Young:2019yug}%
  \BibitemOpen
  \bibfield  {author} {\bibinfo {author} {\bibfnamefont {S.}~\bibnamefont
  {Young}}, \bibinfo {author} {\bibfnamefont {I.}~\bibnamefont {Musco}}, \ and\
  \bibinfo {author} {\bibfnamefont {C.~T.}\ \bibnamefont {Byrnes}},\ }\href
  {\doibase 10.1088/1475-7516/2019/11/012} {\bibfield  {journal} {\bibinfo
  {journal} {JCAP}\ }\textbf {\bibinfo {volume} {11}},\ \bibinfo {pages} {012}
  (\bibinfo {year} {2019})},\ \Eprint {http://arxiv.org/abs/1904.00984}
  {arXiv:1904.00984 [astro-ph.CO]} \BibitemShut {NoStop}%
\bibitem [{\citenamefont {Parker}(1980{\natexlab{a}})}]{Parker:1980hlc}%
  \BibitemOpen
  \bibfield  {author} {\bibinfo {author} {\bibfnamefont {L.}~\bibnamefont
  {Parker}},\ }\href {\doibase 10.1103/PhysRevLett.44.1559} {\bibfield
  {journal} {\bibinfo  {journal} {Phys. Rev. Lett.}\ }\textbf {\bibinfo
  {volume} {44}},\ \bibinfo {pages} {1559} (\bibinfo {year}
  {1980}{\natexlab{a}})}\BibitemShut {NoStop}%
\bibitem [{\citenamefont {Parker}(1980{\natexlab{b}})}]{Parker:1980kw}%
  \BibitemOpen
  \bibfield  {author} {\bibinfo {author} {\bibfnamefont {L.}~\bibnamefont
  {Parker}},\ }\href {\doibase 10.1103/PhysRevD.22.1922} {\bibfield  {journal}
  {\bibinfo  {journal} {Phys. Rev. D}\ }\textbf {\bibinfo {volume} {22}},\
  \bibinfo {pages} {1922} (\bibinfo {year} {1980}{\natexlab{b}})}\BibitemShut
  {NoStop}%
\bibitem [{\citenamefont {Parker}\ and\ \citenamefont
  {Pimentel}(1982)}]{Parker:1982nk}%
  \BibitemOpen
  \bibfield  {author} {\bibinfo {author} {\bibfnamefont {L.}~\bibnamefont
  {Parker}}\ and\ \bibinfo {author} {\bibfnamefont {L.~O.}\ \bibnamefont
  {Pimentel}},\ }\href {\doibase 10.1103/PhysRevD.25.3180} {\bibfield
  {journal} {\bibinfo  {journal} {Phys. Rev. D}\ }\textbf {\bibinfo {volume}
  {25}},\ \bibinfo {pages} {3180} (\bibinfo {year} {1982})}\BibitemShut
  {NoStop}%
\bibitem [{\citenamefont {Shankaranarayanan}\ and\ \citenamefont
  {Johnson}(2022)}]{Shankaranarayanan:2022wbx}%
  \BibitemOpen
  \bibfield  {author} {\bibinfo {author} {\bibfnamefont {S.}~\bibnamefont
  {Shankaranarayanan}}\ and\ \bibinfo {author} {\bibfnamefont {J.~P.}\
  \bibnamefont {Johnson}},\ }\href {\doibase 10.1007/s10714-022-02927-2}
  {\bibfield  {journal} {\bibinfo  {journal} {Gen. Rel. Grav.}\ }\textbf
  {\bibinfo {volume} {54}},\ \bibinfo {pages} {44} (\bibinfo {year} {2022})},\
  \Eprint {http://arxiv.org/abs/2204.06533} {arXiv:2204.06533 [gr-qc]}
  \BibitemShut {NoStop}%
\bibitem [{\citenamefont {Mandal}\ and\ \citenamefont
  {Shankaranarayanan}(2025)}]{Mandal:2025xuc}%
  \BibitemOpen
  \bibfield  {author} {\bibinfo {author} {\bibfnamefont {S.}~\bibnamefont
  {Mandal}}\ and\ \bibinfo {author} {\bibfnamefont {S.}~\bibnamefont
  {Shankaranarayanan}},\ }\href@noop {} {\bibfield  {journal} {\bibinfo
  {journal} {arXiv}\ } (\bibinfo {year} {2025})},\ \Eprint
  {http://arxiv.org/abs/2502.07437} {2502.07437 [gr-qc]} \BibitemShut {NoStop}%
\bibitem [{\citenamefont {Feynman}\ \emph {et~al.}(2006)\citenamefont
  {Feynman}, \citenamefont {Leighton},\ and\ \citenamefont
  {Sands}}]{Feynman2006Vol3}%
  \BibitemOpen
  \bibfield  {author} {\bibinfo {author} {\bibfnamefont {R.~P.}\ \bibnamefont
  {Feynman}}, \bibinfo {author} {\bibfnamefont {R.~B.}\ \bibnamefont
  {Leighton}}, \ and\ \bibinfo {author} {\bibfnamefont {M.}~\bibnamefont
  {Sands}},\ }\href@noop {} {\emph {\bibinfo {title} {The Feynman Lectures on
  Physics, Vol. 3: Quantum Mechanics}}},\ \bibinfo {edition} {new millennium
  edition}\ ed.\ (\bibinfo  {publisher} {Pearson/Addison-Wesley},\ \bibinfo
  {address} {San Francisco, CA},\ \bibinfo {year} {2006})\BibitemShut {NoStop}%
\bibitem [{\citenamefont {Manasse}\ and\ \citenamefont
  {Misner}(1963)}]{Manasse:1963zz}%
  \BibitemOpen
  \bibfield  {author} {\bibinfo {author} {\bibfnamefont {F.~K.}\ \bibnamefont
  {Manasse}}\ and\ \bibinfo {author} {\bibfnamefont {C.~W.}\ \bibnamefont
  {Misner}},\ }\href {\doibase 10.1063/1.1724316} {\bibfield  {journal}
  {\bibinfo  {journal} {J. Math. Phys.}\ }\textbf {\bibinfo {volume} {4}},\
  \bibinfo {pages} {735} (\bibinfo {year} {1963})}\BibitemShut {NoStop}%
\bibitem [{\citenamefont {Poisson}(2009)}]{Poisson:2009pwt}%
  \BibitemOpen
  \bibfield  {author} {\bibinfo {author} {\bibfnamefont {E.}~\bibnamefont
  {Poisson}},\ }\href {\doibase 10.1017/CBO9780511606601} {\emph {\bibinfo
  {title} {{A Relativist's Toolkit: The Mathematics of Black-Hole
  Mechanics}}}}\ (\bibinfo  {publisher} {Cambridge University Press},\ \bibinfo
  {year} {2009})\BibitemShut {NoStop}%
\bibitem [{\citenamefont {Parvez}\ and\ \citenamefont
  {Shankaranarayanan}(2025)}]{Parvez:2025wtq}%
  \BibitemOpen
  \bibfield  {author} {\bibinfo {author} {\bibfnamefont {T.}~\bibnamefont
  {Parvez}}\ and\ \bibinfo {author} {\bibfnamefont {S.}~\bibnamefont
  {Shankaranarayanan}},\ }\href@noop {} {\  (\bibinfo {year} {2025})},\ \Eprint
  {http://arxiv.org/abs/2511.14047} {arXiv:2511.14047 [gr-qc]} \BibitemShut
  {NoStop}%
\bibitem [{\citenamefont {Osterbrock}\ and\ \citenamefont
  {Ferland}(2006)}]{Osterbrock:2006}%
  \BibitemOpen
  \bibfield  {author} {\bibinfo {author} {\bibfnamefont {D.~E.}\ \bibnamefont
  {Osterbrock}}\ and\ \bibinfo {author} {\bibfnamefont {G.~J.}\ \bibnamefont
  {Ferland}},\ }\href@noop {} {\emph {\bibinfo {title} {{Astrophysics of
  Gaseous Nebulae and Active Galactic Nuclei}}}},\ \bibinfo {edition} {2nd}\
  ed.\ (\bibinfo  {publisher} {University Science Books},\ \bibinfo {year}
  {2006})\BibitemShut {NoStop}%
\bibitem [{\citenamefont {Draine}(2011)}]{Draine:2011}%
  \BibitemOpen
  \bibfield  {author} {\bibinfo {author} {\bibfnamefont {B.~T.}\ \bibnamefont
  {Draine}},\ }\href@noop {} {\emph {\bibinfo {title} {{Physics of the
  Interstellar and Intergalactic Medium}}}}\ (\bibinfo  {publisher} {Princeton
  University Press},\ \bibinfo {year} {2011})\BibitemShut {NoStop}%
\bibitem [{\citenamefont {{Anderson}}\ \emph {et~al.}(2014)\citenamefont
  {{Anderson}}, \citenamefont {{Bania}}, \citenamefont {{Balser}},
  \citenamefont {{Cunningham}}, \citenamefont {{Wenger}}, \citenamefont
  {{Johnstone}},\ and\ \citenamefont {{Armentrout}}}]{Anderson:2014}%
  \BibitemOpen
  \bibfield  {author} {\bibinfo {author} {\bibfnamefont {L.~D.}\ \bibnamefont
  {{Anderson}}}, \bibinfo {author} {\bibfnamefont {T.~M.}\ \bibnamefont
  {{Bania}}}, \bibinfo {author} {\bibfnamefont {D.~S.}\ \bibnamefont
  {{Balser}}}, \bibinfo {author} {\bibfnamefont {V.}~\bibnamefont
  {{Cunningham}}}, \bibinfo {author} {\bibfnamefont {T.~V.}\ \bibnamefont
  {{Wenger}}}, \bibinfo {author} {\bibfnamefont {B.~M.}\ \bibnamefont
  {{Johnstone}}}, \ and\ \bibinfo {author} {\bibfnamefont {W.~P.}\ \bibnamefont
  {{Armentrout}}},\ }\href {\doibase 10.1088/0067-0049/212/1/1} {\bibfield
  {journal} {\bibinfo  {journal} {The Astrophysical Journal Supplement Series}\
  }\textbf {\bibinfo {volume} {212}},\ \bibinfo {eid} {1} (\bibinfo {year}
  {2014})},\ \Eprint {http://arxiv.org/abs/1312.6202} {arXiv:1312.6202
  [astro-ph.GA]} \BibitemShut {NoStop}%
\bibitem [{\citenamefont {{Wenger}}\ \emph {et~al.}(2019)\citenamefont
  {{Wenger}}, \citenamefont {{Dickey}}, \citenamefont {{Jordan}}, \citenamefont
  {{Balser}}, \citenamefont {{Armentrout}}, \citenamefont {{Anderson}},
  \citenamefont {{Bania}}, \citenamefont {{Dawson}}, \citenamefont
  {{McClure-Griffiths}},\ and\ \citenamefont {{Rathborne}}}]{Wenger:2019}%
  \BibitemOpen
  \bibfield  {author} {\bibinfo {author} {\bibfnamefont {T.~V.}\ \bibnamefont
  {{Wenger}}}, \bibinfo {author} {\bibfnamefont {J.~M.}\ \bibnamefont
  {{Dickey}}}, \bibinfo {author} {\bibfnamefont {C.~H.}\ \bibnamefont
  {{Jordan}}}, \bibinfo {author} {\bibfnamefont {D.~S.}\ \bibnamefont
  {{Balser}}}, \bibinfo {author} {\bibfnamefont {W.~P.}\ \bibnamefont
  {{Armentrout}}}, \bibinfo {author} {\bibfnamefont {L.~D.}\ \bibnamefont
  {{Anderson}}}, \bibinfo {author} {\bibfnamefont {T.~M.}\ \bibnamefont
  {{Bania}}}, \bibinfo {author} {\bibfnamefont {J.~R.}\ \bibnamefont
  {{Dawson}}}, \bibinfo {author} {\bibfnamefont {N.~M.}\ \bibnamefont
  {{McClure-Griffiths}}}, \ and\ \bibinfo {author} {\bibfnamefont {J.~M.}\
  \bibnamefont {{Rathborne}}},\ }\href {\doibase 10.3847/1538-4365/aaf9a0}
  {\bibfield  {journal} {\bibinfo  {journal} {The Astrophysical Journal
  Supplement Series}\ }\textbf {\bibinfo {volume} {240}},\ \bibinfo {pages}
  {24} (\bibinfo {year} {2019})}\BibitemShut {NoStop}%
\bibitem [{\citenamefont {Guzman}\ \emph {et~al.}(2017)\citenamefont {Guzman},
  \citenamefont {Badnell}, \citenamefont {Chatzikos}, \citenamefont {van Hoof},
  \citenamefont {Williams},\ and\ \citenamefont {Ferland}}]{Guzman:2017xdg}%
  \BibitemOpen
  \bibfield  {author} {\bibinfo {author} {\bibfnamefont {F.}~\bibnamefont
  {Guzman}}, \bibinfo {author} {\bibfnamefont {N.~R.}\ \bibnamefont {Badnell}},
  \bibinfo {author} {\bibfnamefont {M.}~\bibnamefont {Chatzikos}}, \bibinfo
  {author} {\bibfnamefont {P.~A.~M.}\ \bibnamefont {van Hoof}}, \bibinfo
  {author} {\bibfnamefont {R.~J.~R.}\ \bibnamefont {Williams}}, \ and\ \bibinfo
  {author} {\bibfnamefont {G.~J.}\ \bibnamefont {Ferland}},\ }\href {\doibase
  10.1093/mnras/stx269} {\bibfield  {journal} {\bibinfo  {journal} {Mon. Not.
  Roy. Astron. Soc.}\ }\textbf {\bibinfo {volume} {467}},\ \bibinfo {pages}
  {3944} (\bibinfo {year} {2017})},\ \Eprint {http://arxiv.org/abs/1701.07913}
  {arXiv:1701.07913 [astro-ph.CO]} \BibitemShut {NoStop}%
\bibitem [{\citenamefont {{Liu}}\ \emph {et~al.}(2026)\citenamefont {{Liu}},
  \citenamefont {{Theuns}}, \citenamefont {{Chan}}, \citenamefont
  {{Richings}},\ and\ \citenamefont {{McLeod}}}]{Liu:2026}%
  \BibitemOpen
  \bibfield  {author} {\bibinfo {author} {\bibfnamefont {Y.}~\bibnamefont
  {{Liu}}}, \bibinfo {author} {\bibfnamefont {T.}~\bibnamefont {{Theuns}}},
  \bibinfo {author} {\bibfnamefont {T.~K.}\ \bibnamefont {{Chan}}}, \bibinfo
  {author} {\bibfnamefont {A.~J.}\ \bibnamefont {{Richings}}}, \ and\ \bibinfo
  {author} {\bibfnamefont {A.~F.}\ \bibnamefont {{McLeod}}},\ }\href {\doibase
  10.1093/mnras/stag429} {\bibfield  {journal} {\bibinfo  {journal} {Mon. Not.
  Roy. Astron. Soc.}\ } (\bibinfo {year} {2026}),\ 10.1093/mnras/stag429},\
  \Eprint {http://arxiv.org/abs/2509.20361} {arXiv:2509.20361 [astro-ph.GA]}
  \BibitemShut {NoStop}%
\bibitem [{\citenamefont {{Kulkarni}}(2022)}]{Kulkarni:2022}%
  \BibitemOpen
  \bibfield  {author} {\bibinfo {author} {\bibfnamefont {S.~R.}\ \bibnamefont
  {{Kulkarni}}},\ }\href {\doibase 10.1088/1538-3873/ac689e} {\bibfield
  {journal} {\bibinfo  {journal} {Pub. of the Astron. Soc. of the Pacific}\
  }\textbf {\bibinfo {volume} {134}},\ \bibinfo {eid} {084302} (\bibinfo {year}
  {2022})},\ \Eprint {http://arxiv.org/abs/2107.09585} {arXiv:2107.09585
  [astro-ph.GA]} \BibitemShut {NoStop}%
\bibitem [{\citenamefont {Chatzikos}\ \emph {et~al.}(2023)\citenamefont
  {Chatzikos} \emph {et~al.}}]{Chatzikos:2023bkk}%
  \BibitemOpen
  \bibfield  {author} {\bibinfo {author} {\bibfnamefont {M.}~\bibnamefont
  {Chatzikos}} \emph {et~al.},\ }\href {\doibase
  10.22201/ia.01851101p.2023.59.02.12} {\bibfield  {journal} {\bibinfo
  {journal} {Rev. Mex. Astron. Astrofis.}\ }\textbf {\bibinfo {volume} {59}},\
  \bibinfo {pages} {327} (\bibinfo {year} {2023})},\ \Eprint
  {http://arxiv.org/abs/2308.06396} {arXiv:2308.06396 [astro-ph.GA]}
  \BibitemShut {NoStop}%
\bibitem [{\citenamefont {Dennison}\ \emph {et~al.}(2005)\citenamefont
  {Dennison}, \citenamefont {Turner},\ and\ \citenamefont
  {Minter}}]{Dennison:2005fx}%
  \BibitemOpen
  \bibfield  {author} {\bibinfo {author} {\bibfnamefont {B.}~\bibnamefont
  {Dennison}}, \bibinfo {author} {\bibfnamefont {B.~E.}\ \bibnamefont
  {Turner}}, \ and\ \bibinfo {author} {\bibfnamefont {A.~H.}\ \bibnamefont
  {Minter}},\ }\href {\doibase 10.1086/462402} {\bibfield  {journal} {\bibinfo
  {journal} {Astrophys. J.}\ }\textbf {\bibinfo {volume} {633}},\ \bibinfo
  {pages} {309} (\bibinfo {year} {2005})},\ \Eprint
  {http://arxiv.org/abs/astro-ph/0505456} {arXiv:astro-ph/0505456} \BibitemShut
  {NoStop}%
\bibitem [{\citenamefont {Bethe}\ and\ \citenamefont
  {Salpeter}(1957)}]{Bethe:1957ncq}%
  \BibitemOpen
  \bibfield  {author} {\bibinfo {author} {\bibfnamefont {H.~A.}\ \bibnamefont
  {Bethe}}\ and\ \bibinfo {author} {\bibfnamefont {E.~E.}\ \bibnamefont
  {Salpeter}},\ }\href {\doibase 10.1007/978-3-662-12869-5} {\emph {\bibinfo
  {title} {{Quantum Mechanics of One- and Two-Electron Atoms}}}}\ (\bibinfo
  {year} {1957})\BibitemShut {NoStop}%
\bibitem [{\citenamefont {Cohen-Tannoudji}\ \emph {et~al.}(1977)\citenamefont
  {Cohen-Tannoudji}, \citenamefont {Diu},\ and\ \citenamefont
  {Lalo{\"e}}}]{CohenTannoudji:1977}%
  \BibitemOpen
  \bibfield  {author} {\bibinfo {author} {\bibfnamefont {C.}~\bibnamefont
  {Cohen-Tannoudji}}, \bibinfo {author} {\bibfnamefont {B.}~\bibnamefont
  {Diu}}, \ and\ \bibinfo {author} {\bibfnamefont {F.}~\bibnamefont
  {Lalo{\"e}}},\ }\href@noop {} {\emph {\bibinfo {title} {{Quantum Mechanics,
  Volume 2}}}}\ (\bibinfo  {publisher} {Wiley-VCH},\ \bibinfo {year} {1977})\
  \bibinfo {note} {see Chapter XII and its Complements for the fine and
  hyperfine structure of the hydrogen atom.}\BibitemShut {Stop}%
\bibitem [{\citenamefont {Klipfel}\ and\ \citenamefont
  {Kaiser}(2026)}]{Klipfel:2026aug}%
  \BibitemOpen
  \bibfield  {author} {\bibinfo {author} {\bibfnamefont {A.~P.}\ \bibnamefont
  {Klipfel}}\ and\ \bibinfo {author} {\bibfnamefont {D.~I.}\ \bibnamefont
  {Kaiser}},\ }\href@noop {} {\  (\bibinfo {year} {2026})},\ \Eprint
  {http://arxiv.org/abs/2601.05935} {arXiv:2601.05935 [hep-ph]} \BibitemShut
  {NoStop}%
\bibitem [{\citenamefont {Bondi}(1952)}]{bondi1952MNRAS}%
  \BibitemOpen
  \bibfield  {author} {\bibinfo {author} {\bibfnamefont {H.}~\bibnamefont
  {Bondi}},\ }\href {\doibase 10.1093/mnras/112.2.195} {\bibfield  {journal}
  {\bibinfo  {journal} {MNRAS}\ }\textbf {\bibinfo {volume} {112}},\ \bibinfo
  {pages} {195} (\bibinfo {year} {1952})}\BibitemShut {NoStop}%
\bibitem [{\citenamefont {{Frank}}\ \emph {et~al.}(2002)\citenamefont
  {{Frank}}, \citenamefont {{King}},\ and\ \citenamefont
  {{Raine}}}]{Franketal-Book}%
  \BibitemOpen
  \bibfield  {author} {\bibinfo {author} {\bibfnamefont {J.}~\bibnamefont
  {{Frank}}}, \bibinfo {author} {\bibfnamefont {A.}~\bibnamefont {{King}}}, \
  and\ \bibinfo {author} {\bibfnamefont {D.~J.}\ \bibnamefont {{Raine}}},\
  }\href@noop {} {\emph {\bibinfo {title} {{Accretion Power in Astrophysics:
  Third Edition}}}}\ (\bibinfo {year} {2002})\BibitemShut {NoStop}%
\bibitem [{\citenamefont {Armitage}(2020)}]{Armitage:2020owb}%
  \BibitemOpen
  \bibfield  {author} {\bibinfo {author} {\bibfnamefont {P.~J.}\ \bibnamefont
  {Armitage}},\ }\href {\doibase 10.1093/astrogeo/ataa032} {\bibfield
  {journal} {\bibinfo  {journal} {Astron. Geophys.}\ }\textbf {\bibinfo
  {volume} {61}},\ \bibinfo {pages} {2.40} (\bibinfo {year} {2020})},\ \Eprint
  {http://arxiv.org/abs/2004.00203} {arXiv:2004.00203 [astro-ph.HE]}
  \BibitemShut {NoStop}%
\end{thebibliography}
%

\end{document}